\documentclass[prd,preprint,superscriptaddress,amsmath,amssymb,nofootinbib]{revtex4}
\usepackage{graphicx}
\usepackage{dcolumn}
\usepackage{bm}
\usepackage{amssymb}
\usepackage{amsmath}
\usepackage{epsfig}    
\usepackage{color}
\usepackage{slashed}
\usepackage{hhline}
\usepackage{youngtab}
\usepackage{multirow}

\def\be{\begin{equation}}
\def\ee{\end{equation}}
\newcommand{\bea}{\begin{eqnarray}}
\newcommand{\eea}{\end{eqnarray}}
\newcommand{\nn}{\nonumber}

\begin{document}


\title{Quark and lepton model with flavor specific dark matter and muon $g-2$ in modular $A_4$ and hidden $U(1)$ symmetries}

\author{Takaaki Nomura}
\email{nomura@scu.edu.cn}
\affiliation{College of Physics, Sichuan University, Chengdu 610065, China}

\author{Hiroshi Okada}
\email{hiroshi.okada@apctp.org}
\affiliation{Asia Pacific Center for Theoretical Physics (APCTP) - Headquarters San 31, Hyoja-dong,
Nam-gu, Pohang 790-784, Korea}
\affiliation{Department of Physics, Pohang University of Science and Technology, Pohang 37673, Republic of Korea}

\date{\today}

\begin{abstract}
We propose a quark and lepton model explaining their masses, mixings, and CP violating phases, introducing modular $A_4$ and hidden gauged $U(1)$ symmetries. The hidden $U(1)$ brings us heavier Majorana fermions that are requested by chiral anomaly cancellations, and we work on a canonical seesaw scenario due to their neutral particles.
In this framework, we search for favorite parameter space to satisfy both the experimental values and show predictions, applying the $\chi$ square analysis.
Then, we discuss a bosonic dark matter candidate that only annihilates into muon state due to the modular $A_4$ flavor symmetry where we suppose the main interaction of dark matter to be Yukawa terms.
And we study muon anomalous magnetic dipole moment where there are not any constraints of lepton flavor violations thanks to this flavor symmetry. Finally, we show the allowed space to satisfy the observed relic density of dark matter and the muon anomalous magnetic dipole moment.

 \end{abstract}
\maketitle

\section{Introduction}
Flavor physics is one of the attractive issues to resolve several mysteries such as structure of quark and lepton mass matrices, muon anomalous magnetic dipole moment, lepton flavor violations (LFVs), flavor changing neutral currents (FCNCs), and (flavor dependent) dark matter (DM) that are sometimes discussed in indirect detection searches. These are often discussed in a framework of beyond the standard model(SM).
Flavor symmetries are well known as a powerful tool to understand these flavor mysteries and lead us to new insight of new physics.
Especially, a modular flavor symmetry is known as a favorable flavor symmetry that was proposed by F. Feruglio in the paper~\cite{Feruglio:2017spp} and he investigated the lepton sector via a modular $A_4$ group.
Then, a large number of articles are arisen; {\it e.g.,}
the following refs. applying the modular $A_4$ symmetry to lepton and quark sectors~\cite{Criado:2018thu, Kobayashi:2018scp, Okada:2018yrn, Okada:2019uoy, deAnda:2018ecu, Novichkov:2018yse, Nomura:2019yft, Okada:2019mjf, Ding:2019zxk, Nomura:2019lnr, Kobayashi:2019xvz, Zhang:2019ngf, Gui-JunDing:2019wap,
Kobayashi:2019gtp, Nomura:2019xsb, Wang:2019xbo, Okada:2020dmb, Okada:2020rjb, Behera:2020lpd, Nomura:2020opk, Nomura:2020cog, Asaka:2020tmo, Okada:2020ukr, Nagao:2020snm, Okada:2020brs, Kobayashi:2018vbk, Yao:2020qyy, Chen:2021zty, deMedeirosVarzielas:2021pug, Nomura:2021yjb, Ding:2021eva, Nagao:2021rio, Okada:2021aoi, Nomura:2021pld, Liu:2021gwa, Nomura:2022hxs, Otsuka:2022rak, Kang:2022psa, Ishiguro:2022pde,
Nomura:2022boj, Du:2022lij, Kikuchi:2022svo,
Gogoi:2022jwf, Abbas:2022slb, Petcov:2022fjf, Kikuchi:2023jap}.
%
The lepton sector including realization of Leptogenesis is discussed in refs.~\cite{Kobayashi:2018wkl, Asaka:2019vev, Kashav:2021zir, Okada:2021qdf, Dasgupta:2021ggp, Behera:2020sfe, Ding:2022bzs}.
In addition, semi-leptonic flavor changing neutral processes such as $b\to s\ell\bar\ell$, LFVs, muon/electron $g-2$, and EDM are discussed in refs.~\cite{Hutauruk:2020xtk, Kobayashi:2021pav, Kobayashi:2022jvy} and inflation model is argued in ref.~\cite{Gunji:2022xig}.
 As another intriguing aspect, stability of DM is achieved by a remnant symmetry such as $Z_2$ even after the modular symmetry breaking in refs.~\cite{Nomura:2019jxj} and the stability also occurs at three fixed points of modulus $\tau$, that is, $Z_2$ on $\tau=i$ and $Z_3$ on $\tau=\omega,\ i\infty$ that is discuused in ref.~\cite{Kobayashi:2021ajl}.
\footnote{Historically, the DM stability has been discussed by traditional non-Abelian discrete flavor symmetries in refs.~\cite{Lavoura:2012cv, Hirsch:2010ru, Boucenna:2011tj}. Decaying DM models have also been investigated by these traditional symmetries~{\it e.g.}\cite{Kajiyama:2010sb, Haba:2010ag, Kajiyama:2013dba}.}
A flavor specific DM (muon specific case) to explain the Fermi-LAT GeV excess is recently discussed in ref.~\cite{Kim:2023jto} applying the modular $A_4$ symmetry. Here, the bosonic DM interacts with muon only due to the modular flavor symmetry and muon $g-2$ is explained at the same time.

{It is thus interesting to construct a quark and lepton model with modular $A_4$ that realizes predictive quark/lepton sector and contains a DM candidate with flavor specific interactions.
A realization of flavor specific DM interaction would help us to avoid constraints from direct detection experiments since they mainly constrain interactions between DM and 1st generation of fermions (electron, up quark and down quark).  
Moreover we can obtain implications to indirect detection of DM as in the Fermi-LAT discussion in ref.~\cite{Kim:2023jto} by controlling flavor specification in DM interactions.}

In this paper, we constract a model based on modular $A_4$ and hidden gauged $U(1)$ symmetries, 
and explain the quark and lepton masses, mixings, and their CP violating phases with a common modulus $\tau$, and search for favorite space of observables and our input parameters.
Then, we discuss a muon specific DM that is controlled by the modular $A_4$ symmetry, and the DM also contributes to the muon $g-2$. We show the allowed region to satisfy the relic density of DM and muon $g-2$ referring to the latest experimental data.

This paper is organized as follows. In Section~II, we present our model setup and formulate the quark and lepton sector
demonstrating each of allowed space. Then, we also display the common parameter benchmark point of quark and lepton. In Section III, we study the DM that only annihilates into muon pairs assuming Yukawa interaction be dominant,
and muon $g-2$.
After formulating them, we show the allowed space of our input parameters. 
Finally we devote the Section~IV to the summary and conclusion of our results.

\section{Model setup and Constraints}
In this section, we review our model and discuss relevant constraints.
At first, we introduce an alternative hidden gauged $U(1)_H$ symmetry where the SM fields are neutral under it.
{In the hidden sector with hidden $U(1)_H$ we consider $SU(2)$ singlet hidden leptons/quarks in Tab.~\ref{tab:new}.  }
In order to cancel chiral anomalies, we introduce a set of hidden quark/lepton chiral superfields $\bar U,\ U$,  $\bar D,\ D$,
 $\bar E,\ E$,  $\bar N$ and $N$ with $U(1)_H$ charge assignments $0(1)$ for $\bar U(U),\ \bar N (N)$ and 
$0(-1)$ for $\bar D(D),\ \bar E(E)$, respectively, that provide four vector-like fermions after the $U(1)$ symmetry breaking. {Note that charge assignment for these superfields is vector-like under the SM gauge symmetry but chiral under hidden $U(1)_H$}.
Here, $\bar N$ plays a role in arising canonical seesaw mechanism~\cite{Yanagida:1979gs, Magg:1980ut,Lazarides:1980nt,Schechter:1980gr,Cheng:1980qt,Mohapatra:1980yp,Bilenky:1980cx} together with Dirac term $\bar N \nu$ and Majorana term $\bar N\bar N$ due to zero charge under this symmetry, and we suppose the number of family of $\bar N$ to be three. Thus, all the other  hidden quark/lepton superfields have to contain three families to cancel the anomalies. 

In order to have interactions inducing vector-like masses such as $\bar U U \varphi_u$, $\bar N N \varphi_u$, $\bar D D \varphi_d$ and $\bar E E \varphi_d$, we have to introduce SM singlet scalars $\varphi_{u,d}$ with $\pm1$ hidden $U(1)$ charge where these scalars break the hidden $U(1)_H$ symmetry spontaneously. 
$\varphi'_{u}$ is needed to have mass term of $ N N$ where $\varphi'_{d}$ is required to cancel the anomaly
from fermionic supersymmetric partner of $\varphi'_{u}$.
In addition to the hidden symmetry, we introduce the modular $A_4$ symmetry to give predictions for quark and lepton, controlling flavor interactions and assuring the stability of a DM candidate.
At first, we assign $A_4$ triplet for $Q$ and $\bar N, N$ and singlets for the other fields.
The assignment for quark sector is the same as ref.~\cite{Okada:2019uoy}.
While the assignment for lepton sector is achieved so that our DM $\chi_u$ has muon specific Yukawa interaction as
can be seen later, where $\chi_d$ has to be introduced to cancel the anomaly from supersymmetry.
The number of modular weight is also important to control the Yukawa interactions.
For example, we assign (0,odd) for $(\bar U(\bar D), U(D))$ so that our DM cannot couple to these fields.
We summarize all the charge assignments and their fields contents of SM in Tab.~\ref{tab:sm},
and new fields in Tab.~\ref{tab:new}.
\begin{center} 
\begin{table}[tb]
\begin{tabular}{|c||c|c|c|c|c|c||c|c|c|c|c|c|c||c|c||}\hline\hline  
&\multicolumn{6}{c||}{ Leptons} & \multicolumn{7}{c||}{Quarks} & \multicolumn{2}{c||}{Higgs} \\\hline
  & ~$L_{e}$~& ~$L_{\mu}$~ & ~$L_{\tau}$~& ~$\bar e$~& ~$\bar \mu$~ & ~$\bar \tau$~ & ~$Q$~  & ~$\bar u$~& ~$\bar c$~ &~$\bar t$~ & ~$\bar d$~& ~$\bar s$~& ~$\bar b$~ & ~$H_u$~ & ~$H_d$~
  \\\hline 
 $SU(2)_L$ & $\bm{2}$  & $\bm{2}$  & $\bm{2}$ & $\bm{1}$   & $\bm{1}$  & $\bm{1}$ & $\bm{2}$ & $\bm{1}$   & $\bm{1}$ & $\bm{1}$  & $\bm{1}$ & $\bm{1}$  & $\bm{1}$  & $\bm{2}$ & $\bm{2}$   \\\hline 
$U(1)_Y$ & $-\frac12$ & $-\frac12$ & $-\frac12$  & $1$& $1$ & $1$  & $\frac16$  & $-\frac23$   & $-\frac23$ & $-\frac23$ & $\frac13$ & $\frac13$ & $\frac13$ & $-\frac12$ & $\frac12$   \\\hline
 $A_4$ & $1$ & $1'$ & $1''$ & $1''$ & $1''$ & $1'$ & $3$ & $1$ & $1''$ & $1'$   & $1$  & $1''$ & $1'$  & $1$ & $1$   \\\hline
 $-k$ & $0$ & $-4$ & $-4$ & $0$ & $-2$ & $-2$ & $-2$ & $-4$ & $-4$ & $-4$  & $0$& $0$& $0$ & $0$ & $0$    \\\hline
\end{tabular}
\caption{Chiral superfield  contents for quarks, leptons and Higgs
with their charge assignments under $SU(2)_L\times U(1)_Y\times U(1)_{H} \times A_4$ in the SM matter sector, where $-k$ is the number of modular weight.}
\label{tab:sm}
\end{table}
\end{center}

\begin{table}[t!]
\begin{tabular}{|c||c|c|c|c||c|c|c|c|c|c|}\hline\hline  
& ~$\bar U,\ U$~&~ $\bar D,\ D$~  &~ $\bar E,\ E$~ & ~$\bar N,\ N$~ & ~$\chi_u$~ & ~$\varphi_u$~ & ~$\chi_d$~ & ~$\varphi_d$~ & ~$\varphi'_u$~ & ~$\varphi'_d$~ \\\hline\hline 
$SU(2)_L$ & $\bm{1}$  & $\bm{1}$  & $\bm{1}$  & $\bm{1}$     & $\bm{1}$  & $\bm{1}$ & $\bm{1}$  & $\bm{1}$ & $\bm{1}$ & $\bm{1}$    \\\hline 
$U(1)_Y$   & {$-\frac23$, $\frac23$} & {$\frac13$, $-\frac13$} & {$1$, $-1$} & $0$  & $0$  & $0$ & $0$  & $0$ & $0$  & $0$    \\\hline
$U(1)_H$   & $0,\ 1$ & $0,\ -1$ & $0, \ -1$   & $0,\ 1$ & $1$  & $-1$   & $-1$  & $+1$   & $-2$
  & $2$   \\\hline
$A_4$   & $ {\rm singlets}$ & ${\rm singlets}$ & $1$ & $3$  & $1$  & $1$ & $1$  & $1$& $1$  & $1$    \\\hline
$-k$   & $(0,{\rm odd})$ & $(0,{\rm odd})$ & $(-1,0)$   & $-2$    & $-2$  & $-3$   & $
-2$  & $-3$  & $0$  & $0$ \\\hline
\end{tabular}
\caption{ 
Charge assignments of the new chiral superfields
under $SU(3)_C\times SU(2)_L\times U(1)_Y\times U(1)_{H} \times A_4$ associated with the hidden sector, where $-k$ is the number of modular weight.}
\label{tab:new}
\end{table}
%
Then, the  invariant valid superpotentials for quark and lepton under these symmetries are given by
\begin{align}
{\cal W}_q
&= \alpha_u {Y^{(6)}_3} \bar u H_u  Q+
\alpha'_u { Y_{3'}^{(6)}} \bar u H_u  Q+
\beta_u {Y^{(6)}_3} \bar c H_u Q+
\beta'_u {Y_{3'}^{(6)}} \bar c H_u  Q \nonumber\\
& +\gamma_u {Y^{(6)}_3} \bar t H_u Q + \gamma'_u {Y_{3'}^{(6)}} \bar t H_u  Q\\
&+\alpha_d {Y^{(2)}_3} \bar d H_d Q+
\beta_d {Y^{(2)}_3} \bar s H_d Q+
\gamma_d {Y^{(2)}_3} \bar b H_d Q,\\
{\cal W}_\ell
&=  y_e \bar e H_d L_e  +y_\mu \bar\mu H_d L_\mu +y_\tau \bar\tau H_d L_\tau
+y'_\ell \bar e H_d L_\tau
\nn\\
&
+ f_\mu \bar \mu E \chi_u
+  y_{D_1} Y^{(2)}_3\bar N H_u L_{L_e} 
+  y_{D_2} Y^{(6)}_3\bar N H_u L_{L_\mu} 
+  y'_{D_2} Y^{(6)}_{3'}\bar N H_u L_{L_\mu} \nn\\
&
+  y_{D_3} Y^{(6)}_3\bar N H_u L_{L_\tau} 
+  y'_{D_3} Y^{(6)}_{3'}\bar N H_u L_{L_\tau} \nn\\
&
+M_{R_1} [\bar N \bar N]_{1}+M_{R_2} [\bar N \bar N]_{1''}
+M_{R_3} [Y^{(4)}_3 \bar N \bar N]_3\nn\\
&
+y_{L_1}\varphi'_u [N N]_{1}+y_{L_2}\varphi'_u [ N  N]_{1''}
+y_{L_3}\varphi'_u [Y^{(4)}_3 N N]_3
, \label{Eq:yuk}
\end{align}
where the
lower indices $(a,b,i)=1, 2, 3$ are the number of families, and
muon Yukawa coupling $y_\mu$ does not mix each others due to the modular $A_4$ symmetry~\footnote{We can have terms $\bar u U \varphi_u$ and $\bar d D \varphi_d$ which induce small mixing between the SM quarks and hidden quarks.  In this work, we consider mass of hidden quarks are sufficiently heavy and omit discussion of hidden quark sector.}.
$Y^{(i)}_{3,3'}$ are modular Yukawa couplings that are found in ref.~\cite{Feruglio:2017spp} where $i=2,4,6,...$ is modular weight and $A_4$-singlet modular Yukawas with relevant modular weight are implicitly included in free parameters such as $y_{e,\mu,\tau},\ f_\mu, \ M_{R_{1,2}},\ y_{L_{1,2}}$.

\subsection{Quark sector}
The up-type quark mass matrix is written as:
\begin{align}
\begin{aligned}
M_u=v_u\gamma_u
\begin{pmatrix}
\tilde\alpha_u & 0 & 0 \\
0 &\tilde\beta_u & 0\\
0 & 0 &1
\end{pmatrix} \left [
\begin{pmatrix}
Y_1^{(6)} & Y_3^{(6)}& Y_2^{(6)} \\
Y_2^{(6)} & Y_1^{(6)} &  Y_3^{(6)} \\
Y_3^{(6)} &  Y_2^{(6)}&  Y_1^{(6)}
\end{pmatrix}
+ 
\begin{pmatrix}
g_{u1} & 0 & 0 \\
0 &g_{u2} & 0\\
0 & 0 &g_{u3}
\end{pmatrix}
\begin{pmatrix}
Y_1^{'(6)} & Y_3^{'(6)}& Y_2^{'(6)} \\
Y_2^{'(6)} & Y_1^{'(6)} &  Y_3^{'(6)} \\
Y_3^{'(6)} &  Y_2^{'(6)}&  Y_1^{'(6)}
\end{pmatrix}
\right ],
\end{aligned}
\label{matrix6}
\end{align}
where $\tilde\alpha_u=\alpha_u/\gamma_u$, $\tilde\beta_u=\beta_u/\gamma_u$, $g_{u1}=\alpha'_u/\alpha_u$, $g_{u2}=\beta'_u/\beta_u$
and $g_{u3}=\gamma'_u/\gamma_u$ are complex parameters while
 $\alpha_u$,  $\beta_u$ and  $\gamma_u$  are  real.
 On the other hand, the down-type quark mass matrix is given as:  
 \begin{align}
 &\begin{aligned}
 M_d= v_d\gamma_d
 \begin{pmatrix}
\tilde \alpha_d & 0 & 0 \\
 0 &\tilde \beta_d & 0\\
 0 & 0 &1
 \end{pmatrix}
 \begin{pmatrix}
 Y_1 & Y_3& Y_2\\
 Y_2 & Y_1 &  Y_3 \\
 Y_3 &  Y_2&  Y_1
 \end{pmatrix}\,,
 \end{aligned} 
 \label{down}
 \end{align}
 where $\tilde\alpha_d=\alpha_d/\gamma_d$, $\tilde\beta_d=\beta_d/\gamma_d$.

In order to obtain the left-handed quark mixing matrices $V_u,\ V_d$,
we diagonalize $M_u^{\dagger} M_u$ and  $M_d^{\dagger} M_d$, respectively.
Then, the CKM mixing is defined by $V_{\rm CKM}=V^\dag_u V_d$.
We search for allowed regions of our quark sector, showing numerical $\Delta \chi^2$ analysis to fit the sixteen reliable experimental data; six quark masses, nine CKM components, and one quark CP phase, where we assume all observables are Gaussian and randomly scan our input parameters as follows~\cite{Okada:2019uoy}:
\begin{align}
|g_{u1,u2,u3}|\in [10^{-2},10^2].
\end{align}
Notice here that $\tau$ runs over the fundamental region and $\alpha_{u,d},\beta_{u,d},\gamma_{u,d}$ are used to fix the quark masses solving the following three relations:
\begin{align}
&{\rm Tr}[M_{u,d} {M_{u,d}}] = |m_{u,d}|^2 + |m_{c,s}|^2 + |m_{t,b}|^2,\\
&{\rm Det}[M^\dag_{u,d} {M_{u,d}}] = |m_{u,d}|^2  |m_{c,s}|^2  |m_{t,b}|^2,\\
&({\rm Tr}[M^\dag_{u,d} {M_{u,d}}])^2 -{\rm Tr}[(M^\dag_{u,d} {M_{u,d}})^2] =2( |m_{u,d}|^2  |m_{c,s}|^2 + |m_{c,s}|^2  |m_{t,b}|^2+ |m_{u,d}|^2  |m_{t,b}|^2 ).
\end{align}
%
We adopt observed values for quark sector with $\tan\beta=5$ 
\cite{Antusch:2013jca, Bjorkeroth:2015ora}:
\begin{align}
&y_d=(4.81\pm 1.06) \times 10^{-6}, \quad y_s=(9.52\pm 1.03) \times 10^{-5}, \quad y_b=(6.95\pm 0.175) \times 10^{-3}, \\
&y_u=(2.92\pm 1.81) \times 10^{-6}, \quad y_c=(1.43{\pm 0.100}) \times 10^{-3}, \quad y_t=0.534\pm 0.0341,
\label{yukawa5}\\
&\theta_{12}^{\rm CKM}=13.027^\circ\pm 0.0814^\circ ~ , \quad
\theta_{23}^{\rm CKM}=2.054^\circ\pm 0.384^\circ ~ ,  \quad
\theta_{13}^{\rm CKM}=0.1802^\circ\pm 0.0281^\circ~ \label{CKM},\\
&\delta_{CP} \ ({\rm quark})=69.21^\circ\pm 6.19^\circ~ ,
\end{align}
where the errors represent $1\sigma$ interval,
quark masses are defined as $m_q=y_q v_H$ with $v_H=174$ GeV,
and $\theta_{ij}^{\rm CKM}, \delta_{CP}$ are the PDG notation~\cite{ParticleDataGroup:2018ovx}; {we write $\delta_Q \equiv \delta_{CP} \ ({\rm quark})$ in later part to distinguish it from CP violating phase in neutrino sector}.

\begin{figure}[tb]
\begin{center}
\includegraphics[width=77.0mm]{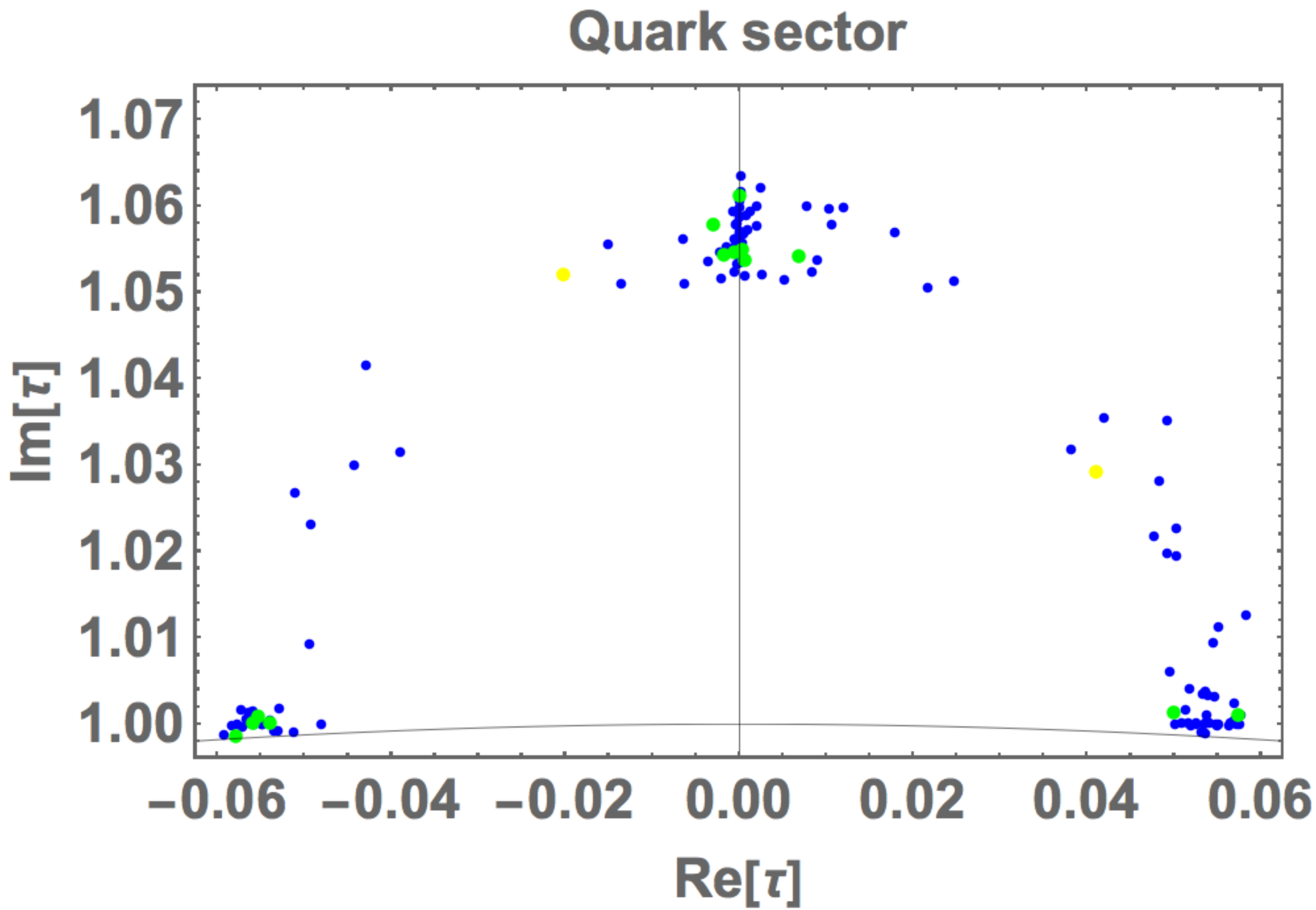}
\caption{Allowed region of real and imaginary part of $\tau$ in fundamental region.
The blue, green, and yellow points correspond to $\sigma\le1$, $1<\sigma\le2$, and $2<\sigma\le3$ interval, respectively in $\Delta\chi^2$ analysis. }
  \label{fig:tauQ}
\end{center}\end{figure}
In Fig.~\ref{fig:tauQ}, we plot the allowed region of real and imaginary part of $\tau$ in fundamental region, where
the blue, green, and yellow points correspond to $\sigma\le1$, $1<\sigma\le2$, and $2<\sigma\le3$ interval, respectively in $\Delta\chi^2$ analysis.
The allowed space is located at nearby $0.05\lesssim|\tau|\lesssim0.06$ which is favored by the analysis in ref.~\cite{Okada:2019uoy}.

\begin{figure}[tb]
\begin{center}
\includegraphics[width=77.0mm]{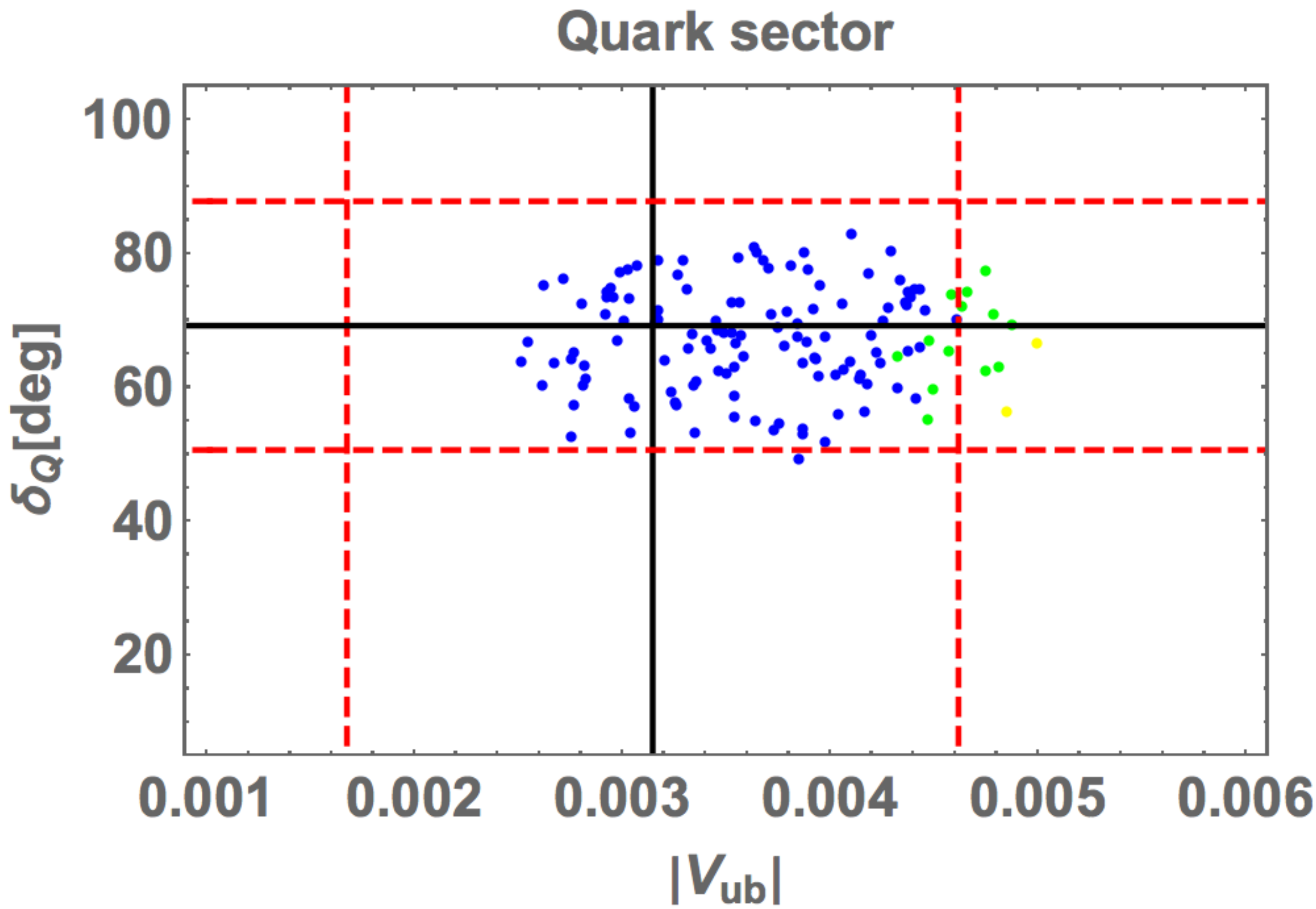}
\caption{The correlation between $|V_{ub}|$ and $\delta_{\rm Q}$ in the quark sector, where the color legends are the same as the ones in Fig.~\ref{fig:tauQ}. The black solid lines represent the best fit values and the red dotted lines show the 3$\sigma$ lines in the experiments.}
  \label{fig:vub-dQ}
\end{center}\end{figure}
In Fig.~\ref{fig:vub-dQ}, we plot the correlation between $|V_{ub}|$ and $\delta_{\rm Q}$ in the quark sector, where the color legends are the same as the ones in Fig.~\ref{fig:tauQ}. The black solid lines represent the best fit values and the red dotted lines show the 3$\sigma$ lines in the experiments.
While the allowed region of $\delta_{\rm Q}$ runs over whole the region within 3$\sigma$ interval, the allowed region of $|V_{ub}|$ tends to be larger value; $0.0025\lesssim|V_{ub}|\lesssim0.005$.

\begin{figure}[tb]
\begin{center}
\includegraphics[width=70.0mm]{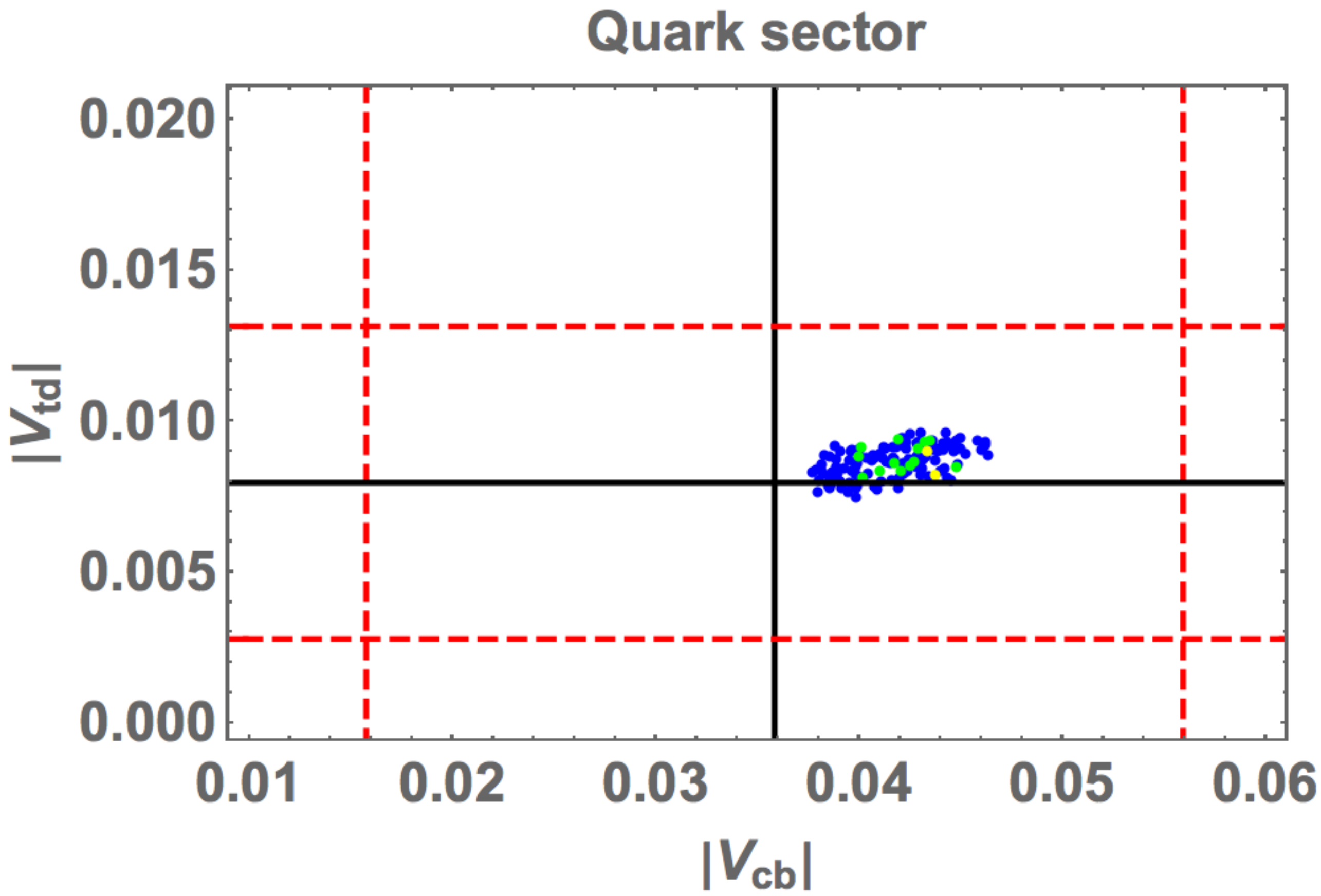} 
\caption{The correlation between $|V_{cb}|$ and $|V_{td}|$ in the quark sector, where the color legends and lines are the same as the ones in Fig.~\ref{fig:vub-dQ}.}
  \label{fig:vcb-vtd}
\end{center}\end{figure}
In Fig.~\ref{fig:vcb-vtd}, we plot the correlation between $|V_{cb}|$ and $|V_{td}|$ in the quark sector, where the color legends and lines are the same as the ones in Fig.~\ref{fig:vub-dQ}.
Both of the allowed regions are localized at $0.038\lesssim|V_{cb}|\lesssim0.047$
and $0.0075\lesssim|V_{td}|\lesssim0.0095$.
In Tabs.~\ref{u-para} and \ref{d-para}, we show our allowed space for the model parameters of each the up and down quark sector. The tendency for up quark sector would be hierarchal, on the other hand the one for down quark would be milder than the up quark one. These features would come from the mass hierarchy for the difference between up-quark and down-quark sector. 

\begin{table}[h!]
	\centering
	\begin{tabular}{|c|c|c|c|c|c|} \hline 
		\rule[14pt]{0pt}{0pt}
		$|g_{u1}|$
		&$|g_{u2}|$ 
		&$|g_{u3}|$
		&$\tilde\alpha_{u}$ 
		&$\tilde\beta_{u}$
		&$\gamma_{u}$
		\rule[14pt]{0pt}{0pt}	 \\   		
		\hline \hline 
		$[0.071,\,62]$ &$[0.014,\,50]$ & $[0.012,\,0.23]$
		&$[2.6\times10^{-5},\,330]$&$[9.2\times10^{-7},\,8.6]$ &  $[4.6\times10^{-5},\,0.27]$ 
		\rule[14pt]{0pt}{0pt}	\\
		\hline
	\end{tabular}
	\caption{Model parameter ranges consistent with the observed values in the up quark sector, where arguments of $g_{u1,u2,u3}$ run over $0-2\pi$.}
	\label{u-para}
\end{table}

\begin{table}[h!]
	\centering
	\begin{tabular}{|c|c|c|} \hline 
		\rule[14pt]{0pt}{0pt}
	        $\tilde\alpha_d$
	       & $\tilde\beta_{d}$
              & $\gamma_d$
		\rule[14pt]{0pt}{0pt}	 \\   		
		\hline \hline 
		 $[0.0033,\,0.015]$ & $[0.0036,\,0.015]$ & $[0.026,\,0.029]$
		\rule[14pt]{0pt}{0pt}	\\
		\hline
	\end{tabular}
	\caption{Output parameter ranges consistent with the observed values in the down quark sector.}
	\label{d-para}
\end{table}

\subsection{Lepton sector}
After the electroweak spontaneous symmetry breaking,  the charged-lepton mass matrix is given by
\begin{align}
m_\ell&= \frac {v_d }{\sqrt{2}}
 \left[\begin{array}{ccc}
y_e &0 & y'_\ell \\ 
0 &y_\mu & 0 \\ 
0 &0 & y_\tau \\ 
 \end{array}\right], 
\end{align}
where $\langle H_d\rangle\equiv [v_d/\sqrt2,0]^T$.
Then the charged-lepton mass eigenvalues are found as ${\rm  diag}( |m_e|^2, |m_\mu|^2, |m_\tau|^2)\equiv V_{e_L}^\dag m^\dag_\ell m_\ell V_{e_L}$.
Similar to the way in quark sector, we fix three input parameters $y_e,y_\tau,y'_\ell$ inserting the observed charged-lepton masses and $V_{e_L}$:
\begin{align}
&{\rm Tr}[m_\ell {m_\ell}^\dag] = |m_e|^2 + |m_\mu|^2 + |m_\tau|^2,\\
&{\rm Det}[m_\ell {m_\ell}^\dag] = |m_e|^2  |m_\mu|^2  |m_\tau|^2,\\
&({\rm Tr}[m_\ell {m_\ell}^\dag])^2 -{\rm Tr}[(m_\ell {m_\ell}^\dag)^2] =2( |m_e|^2  |m_\mu|^2 + |m_\mu|^2  |m_\tau|^2+ |m_e|^2  |m_\tau|^2 ).
\end{align}
Notice here that $y_\mu$ is determined by the mass of muon $y_\mu= \sqrt2 m_\mu/v_d$, and $y_e\approx1.18$, $y'_\ell\approx0.589$, and $y_\tau\approx0.000581$ with the observed charged-lepton masses.
We apply the experimental values summarized in PDG for the charged-lepton masses~\cite{ParticleDataGroup:2018ovx}.

The Dirac mass matrix under basis of $\bar N m_D \nu$ is given by
\begin{align}
m_D &= \frac{y_{D_1} v_d}{\sqrt2}
 \left[\begin{array}{ccc}
y_1 & y_3^{(6)}+\epsilon_2 y'^{(6)}_3 & y_2^{(6)}+\epsilon_3 y'^{(6)}_2 \\ 
y_3 & y_2^{(6)}+\epsilon_2 y'^{(6)}_2 & y_1^{(6)}+\epsilon_3 y'^{(6)}_1 \\ 
y_2 & y_1^{(6)}+\epsilon_2 y'^{(6)}_1 & y_3^{(6)}+\epsilon_3 y'^{(6)}_3 \\ 
\end{array}\right]
 \left[\begin{array}{ccc}
1 & 0 & 0 \\ 
0 & \tilde y_{D_2} & 0 \\ 
0 & 0 & \tilde y_{D_3} \\ 
\end{array}\right]
\equiv \frac{y_{D_1} v_d}{\sqrt2} \tilde m_D ,
\end{align}
where $\tilde y_{D_i}\equiv y_{D_i}/y_{D_1}$ and $\epsilon_i\equiv y_{D_i}/y'_{D_i}$ (i=2,3).
The heavier Majorana mass matrix is given by
\begin{align}
M_N &= M_{R_3}
\left(
 \left[\begin{array}{ccc}
2 y^{(4)}_1 & -y^{(4)}_3 & -y^{(4)}_2 \\ 
- y^{(4)}_3 & 2y^{(4)}_2 & -y^{(4)}_1 \\ 
- y^{(4)}_2 & -y^{(4)}_1 & 2y^{(4)}_3 \\ 
\end{array}\right]
+ \epsilon_{M_1}
 \left[\begin{array}{ccc}
1 & 0 & 0 \\ 
0 & 0 & 1 \\ 
0 & 1 & 0 \\ 
\end{array}\right]
+ \epsilon_{M_2}
 \left[\begin{array}{ccc}
0 & 0 & 1 \\ 
0 & 1 & 0 \\ 
1 & 0 & 0 \\ 
\end{array}\right]
\right)
=
M_{R_3} {\tilde M},
\end{align}
where $\epsilon_{M_i}\equiv M_{R_i}/M_{R_3}$ (i=1,2).
The heavy Majorana mass matrix is diagonalized by  a unitary matrix $V_N$ as follows: $D_N\equiv V_N M_N V_N^T$,
where $N^c\equiv \psi^cV_N^T$, $\psi^c$ being mass eigenstate.

If all the bosons have nonzero VEVs, the neutrino mass matrix is generated via canonical seesaw at tree-level as follows:
\begin{align}
m_\nu 
= \frac{y_{D_1}^2 v_{d}^2}{2 M_{R_3}}
\tilde m_D^T \tilde M^{-1}\tilde m_D \equiv \kappa \tilde m_\nu,
\end{align}
where $\kappa\equiv    \frac{y_{D_1}^2 v_{d}^2}{2 M_{R_3}}$.
$m_\nu$ is diagonalzied by a unitary matrix $V_{\nu}$; $D_\nu=|\kappa| \tilde D_\nu= V_{\nu}^T m_\nu V_{\nu}=|\kappa| V_{\nu}^T \tilde m_\nu V_{\nu}$.
Then $|\kappa|$ is determined by
\begin{align}
(\mathrm{NH}):\  |\kappa|^2= \frac{|\Delta m_{\rm atm}^2|}{\tilde D_{\nu_3}^2-\tilde D_{\nu_1}^2},
\quad
(\mathrm{IH}):\  |\kappa|^2= \frac{|\Delta m_{\rm atm}^2|}{\tilde D_{\nu_2}^2-\tilde D_{\nu_3}^2},
 \end{align}
where $\Delta m_{\rm atm}^2$ is atmospheric neutrino mass-squared splitting, and NH and IH respectively represent the normal hierarchy and the inverted hierarchy. 
Subsequently, the solar mass-squared splitting can be written in terms of $|\kappa|$ as follows:
\begin{align}
\Delta m_{\rm sol}^2=  |\kappa|^2 ({\tilde D_{\nu_2}^2-\tilde D_{\nu_1}^2}),
 \end{align}
 which can be compared to the observed value.
 %
The observed mixing matrix is defined by $U=V^\dag_L V_\nu$~\cite{Maki:1962mu}, where
it is parametrized by three mixing angles $\theta_{ij} (i,j=1,2,3; i < j)$, one CP violating Dirac phase $\delta_{CP}$,
and two Majorana phases $\alpha_{21},\alpha_{31}$ as follows:
\begin{equation}
U = 
\begin{pmatrix} c_{12} c_{13} & s_{12} c_{13} & s_{13} e^{-i \delta_{CP}} \\ 
-s_{12} c_{23} - c_{12} s_{23} s_{13} e^{i \delta_{CP}} & c_{12} c_{23} - s_{12} s_{23} s_{13} e^{i \delta_{CP}} & s_{23} c_{13} \\
s_{12} s_{23} - c_{12} c_{23} s_{13} e^{i \delta_{CP}} & -c_{12} s_{23} - s_{12} c_{23} s_{13} e^{i \delta_{CP}} & c_{23} c_{13} 
\end{pmatrix}
\begin{pmatrix} 1 & 0 & 0 \\ 0 & e^{i \frac{\alpha_{21}}{2}} & 0 \\ 0 & 0 &  e^{i \frac{\alpha_{31}}{2}} \end{pmatrix}.
\end{equation}
Here, $c_{ij}$ and $s_{ij}$ stand for $\cos \theta_{ij}$ and $\sin \theta_{ij}$ ($i,j=1-3$), respectively. 
Then, each of the mixings is given in terms of the component of $U$ as follows:
\begin{align}
\sin^2\theta_{13}=|U_{e3}|^2,\quad 
\sin^2\theta_{23}=\frac{|U_{\mu3}|^2}{1-|U_{e3}|^2},\quad 
\sin^2\theta_{12}=\frac{|U_{e2}|^2}{1-|U_{e3}|^2}.
\end{align}
The Dirac phase  $\delta_{CP}$ is given by computing  the Jarlskog invariant as follows:
\begin{align}
\sin \delta_{CP} &= \frac{\text{Im} [U_{e1} U_{\mu 2} U_{e 2}^* U_{\mu 1}^*] }{s_{23} c_{23} s_{12} c_{12} s_{13} c^2_{13}} ,\quad
\cos \delta_{CP} = -\frac{|U_{\tau1}|^2 -s^2_{12}s^2_{23}-c^2_{12}c^2_{23}s^2_{13}}{2 c_{12} s_{12} c_{23} s_{23}s_{13}} ,
\end{align}
where $\delta_{CP}$ be subtracted from $\pi$ if $\cos \delta_{CP}$ is negative.
Majorana phase $\alpha_{21},\ \alpha_{31}$ are found as
\begin{align}
&
\sin \left( \frac{\alpha_{21}}{2} \right) = \frac{\text{Im}[U^*_{e1} U_{e2}] }{ c_{12} s_{12} c_{13}^2} ,\quad
  \cos \left( \frac{\alpha_{21}}{2} \right)= \frac{\text{Re}[U^*_{e1} U_{e2}] }{ c_{12} s_{12} c_{13}^2}, \
%
,\\
&
 \sin \left(\frac{\alpha_{31}}{2}  - \delta_{CP} \right)=\frac{\text{Im}[U^*_{e1} U_{e3}] }{c_{12} s_{13} c_{13}},
\quad 
 \cos \left(\frac{\alpha_{31}}{2}  - \delta_{CP} \right)=\frac{\text{Re}[U^*_{e1} U_{e3}] }{c_{12} s_{13} c_{13}},
\end{align}
where $\alpha_{21}/2,\ \alpha_{31}/2-\delta_{CP}$
are subtracted from $\pi$, when $ \cos \left( \frac{\alpha_{21}}{2} \right),\  \cos \left(\frac{\alpha_{31}}{2}  - \delta_{CP} \right)$ are negative.
In addition, the effective mass for the neutrinoless double beta decay is given by
\begin{align}
\langle m_{ee}\rangle=|\kappa||\tilde D_{\nu_1} \cos^2\theta_{12} \cos^2\theta_{13}+\tilde D_{\nu_2} \sin^2\theta_{12} \cos^2\theta_{13}e^{i\alpha_{21}}+\tilde D_{\nu_3} \sin^2\theta_{13}e^{-2i\delta_{CP}}|,
\end{align}
where its observed value could be measured by KamLAND-Zen in future~\cite{KamLAND-Zen:2016pfg}.

Next, we demonstrate the allowed space, applying $\Delta\chi$ square analysis to satisfy the current neutrino oscillation data. We randomly select within the ranges of six input dimensionless parameters $\tilde y_{D_{2,3}},\ |\epsilon_{2,3}|,\ |\epsilon_{M_{1,2}}|$ as $[10^{-3}-10^3]$.~\footnote{$\tilde y_{D_{2,3}}$ can be real after rephasing, while the others are complex.}
{\it We work on the region of $|\tau|\le0.1$ considering the allowed region of quark sector.}
Then, we take into consideration of five reliable measured neutrino oscillation data ($\Delta m^2_{\rm atm},\ \Delta m^2_{\rm sol},\ s^2_{13},\ s^2_{23},\ s^2_{12}$) in Nufit 5.2~\cite{Esteban:2020cvm}, while $\delta_{CP}$ is treated as an output parameter due to large ambiguity of experimental result in $3\sigma$ interval.
{\it In case of IH, we have found the allowed region of $|{\rm Re}[\tau]|=[0.07-0.09]$ and ${\rm Im}[\tau]=[2.2-3.1]$ which is out of the region allowed by quark sector. Thus, we concentrate on NH only.}

\begin{figure}[tb]
\includegraphics[width=77.0mm]{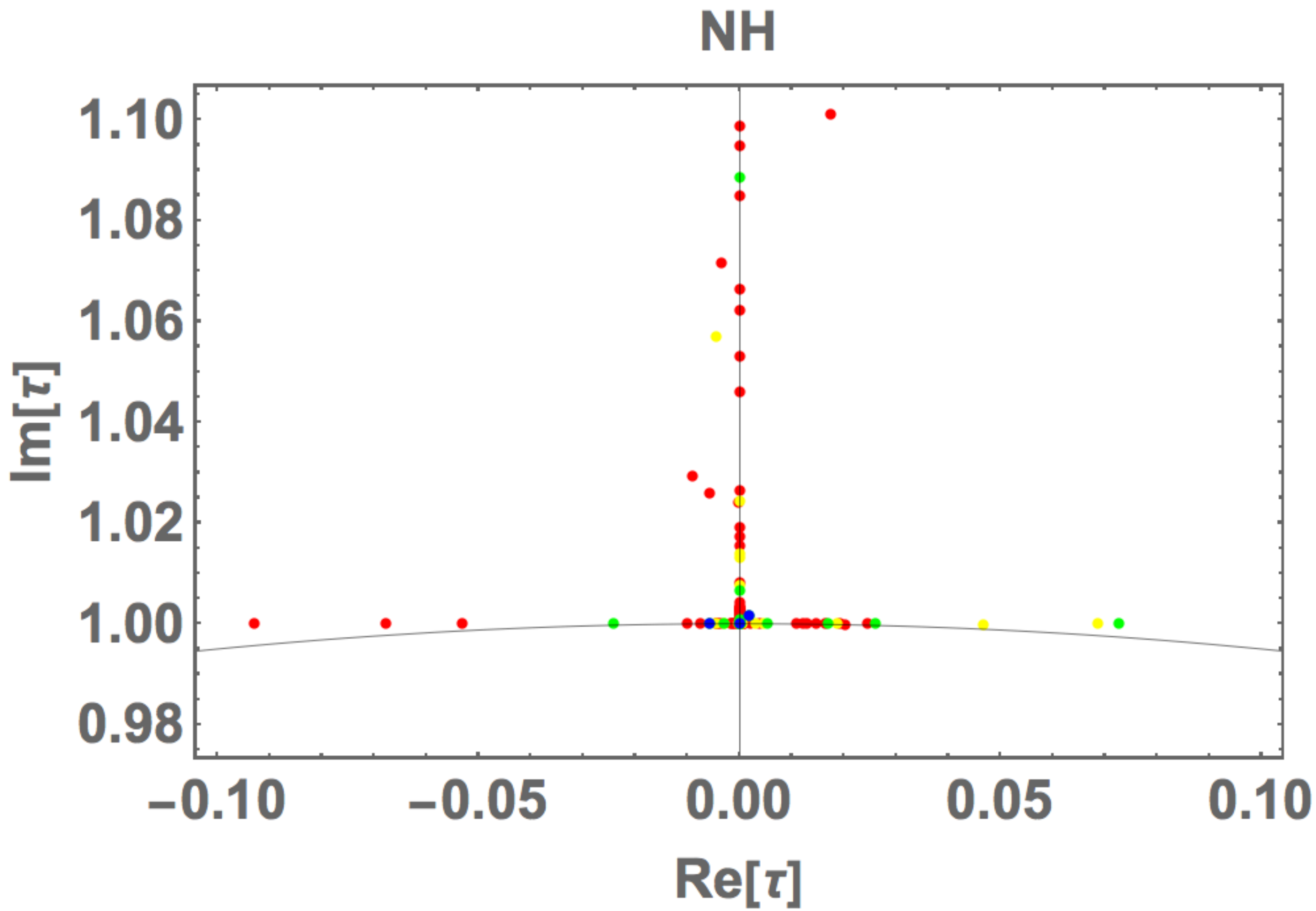}
 \caption{Allowed region between real and imaginary part
of $\tau$, where each of color represents ${\rm blue} \le1\sigma$, $1\sigma< {\rm green}\le 2\sigma$, $2\sigma< {\rm yellow}\le 3\sigma$, $3\sigma<{\rm red}\le5\sigma$.}
 \label{fig:tau}
\end{figure}
In Fig.~\ref{fig:tau}, we show the allowed region between real and imaginary part
of $\tau$, where each of color represents ${\rm blue} \le1\sigma$, $1\sigma< {\rm green}\le 2\sigma$, $2\sigma< {\rm yellow}\le 3\sigma$, $3\sigma<{\rm red}\le5\sigma$.
The figure tells us that the allowed points are localized at nearby the fundamental lines $\sqrt{1-({\rm Re}[\tau])^2}$ and ${\rm Re}[\tau]=0$.

\begin{figure}[tb]
\includegraphics[width=77.0mm]{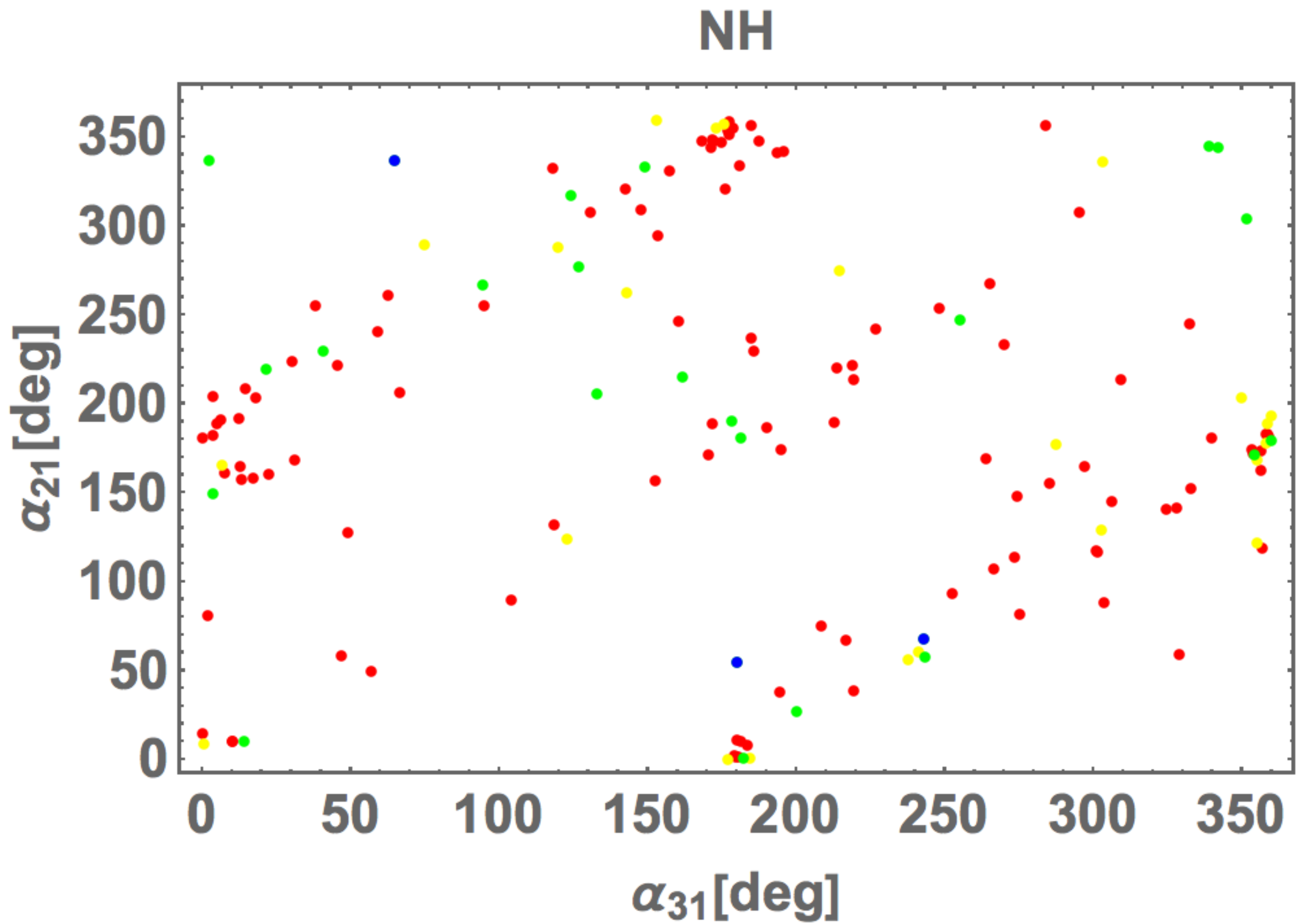}
 \caption{Allowed region between Majorana phases, where color legends are the same as the one in Fig.~\ref{fig:tau}.}
 \label{fig:majo}
\end{figure}
In Fig.~\ref{fig:majo}, we show the allowed region between Majorana phases, where color legends are the same as the one in Fig.~\ref{fig:tau}.
Even though there exist a weak linear correlation between $\alpha_{21}$ and $\alpha_{31}$, any values would be allowed.
Although we do not show the figure related to Dirac CP phase, we have checked $\delta_{CP}$ runs over whole the region.

\begin{figure}[tb]
   \includegraphics[width=77.0mm]{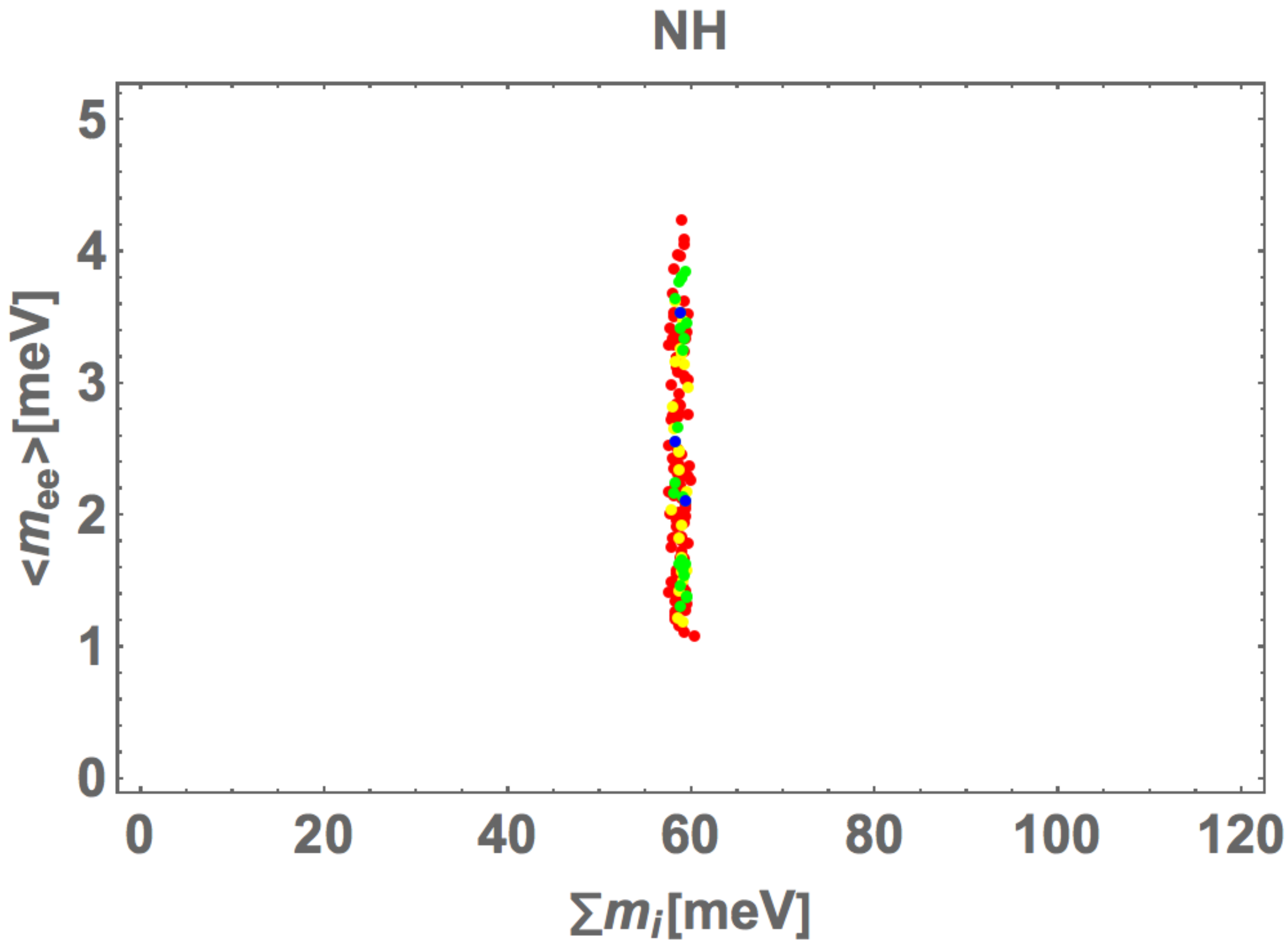}
 \caption{Allowed region between sum of masses and the neutrinoless double beta decay, where color legends are the same as the one in Fig.~\ref{fig:tau}.}
 \label{fig:masses}
\end{figure}
In Fig.~\ref{fig:masses}, we show the allowed region between sum of neutrino masses and the double beta decay, where color legends are the same as the one in Fig.~\ref{fig:tau}.
The figure clearly show that sum of masses is localized at nearby $60$ meV, while $\langle m_{ee}\rangle=[1-4.2]$ meV.
In Tab.~\ref{lep-para}, we show our allowed space for input parameters of neutrino sector.
The parameters $\epsilon_{M_{1,2}}, \epsilon_2$ almost run whole the range of of [$10^{-3}-10^3$], while the other parameters are rather restricted by explaining the neutrino oscillation data.

\begin{table}[h!]
	\centering
	\begin{tabular}{|c|c|c|c|c|c|} \hline 
		\rule[14pt]{0pt}{0pt}
		$|\epsilon_{M_1}|$
		&$|\epsilon_{M_2}|$ 
		&$\tilde y_{D_2}$
		&$\tilde y_{D_3}$ 
		&$|\epsilon_{2}|$
		&$|\epsilon_{3}|$
		\rule[14pt]{0pt}{0pt}	 \\   		
		\hline \hline 
		$[0.00626,\,867]$ &$[0.00278,\,896]$ & $[0.00124,\,1.828]$
		&$[0.0425,\,0.975]$&$[0.0164,\,947]$ &  $[0.00219,\,2.67]$ 
		\rule[14pt]{0pt}{0pt}	\\
		\hline
	\end{tabular}
	\caption{Input parameter ranges consistent with the observed values in the lepton sector, where arguments of $\epsilon_{M_{1,2}}, \ \epsilon_{2,3}$ run over $0-2\pi$.}
	\label{lep-para}
\end{table}

\subsection{Common region of quark and lepton}
\label{sec:crql}
Now that we have searched for each the allowed region of quark and lepton sector,
we investigate the common region of $\tau$ satisfying both sectors.
\begin{figure}[tb]
\includegraphics[width=88.0mm]{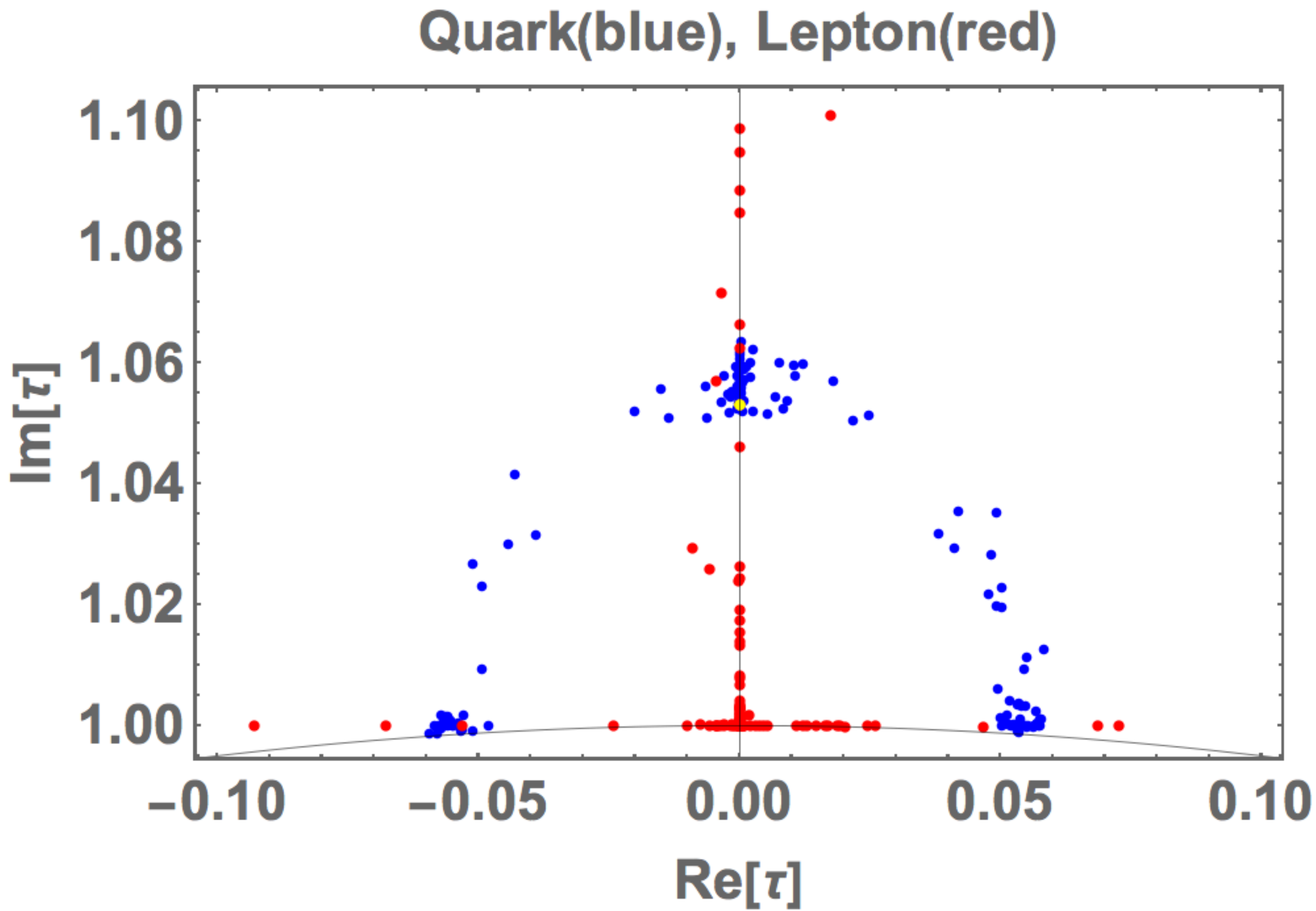}
 \caption{Allowed region between real and imaginary part
of $\tau$ in case of quark(blue) and lepton(red), where quark sector is within $3\sigma$ and lepton sector within $5\sigma$. We select the benchmark point as the yellow dot satisfying both the sectors.}
 \label{fig:tau_QL}
\end{figure}
In Fig.~\ref{fig:tau_QL}, we show the allowed region between real and imaginary part
of $\tau$ in case of quark(blue) and lepton(red), where quark sector is within $3\sigma$ and lepton sector within $5\sigma$. One finds that there exist overlap space for both sector. 
Here, we select the benchmark point (BP) as a yellow dot $\tau\approx1.053i$.
In Tab.~\ref{BP-q}, we show the BP for quark sector.
In Tab.~\ref{BP-l}, we show the BP for lepton sector.

 \begin{table}[h!]
 		\centering
 	\begin{tabular}{|c|c|} \hline 
 		\rule[14pt]{0pt}{0pt}	
 		$\tau$& $1.053 \, i$  \\ 
 		\rule[14pt]{0pt}{0pt}
 		$g_{u1}$ &$-20.4551 - 3.89777 i$ \\
 		\rule[14pt]{0pt}{0pt}
 		$g_{u2}$ & $ 0.0422226 + 0.308798 i$ \\
 		\rule[14pt]{0pt}{0pt}
 		$g_{u3}$ & $0.0130486 + 0.0159581 i$  \\
 		\rule[14pt]{0pt}{0pt}
 	$\hat\alpha_u$ & $0.713292$  \\
 		\rule[14pt]{0pt}{0pt}
 	$\hat\beta_u$ & $ 0.000426951$  \\
 		\rule[14pt]{0pt}{0pt}
	$\gamma_u$ & $ 0.023187$  \\
 		\rule[14pt]{0pt}{0pt}
 	$\hat\alpha_d$ & $0.00392545$  \\
 		\rule[14pt]{0pt}{0pt}
 		$\hat\beta_d$ & $0.0152835$ \\
 		\rule[14pt]{0pt}{0pt}
	$\gamma_d$ & $ 0.028698$  \\
 		\rule[14pt]{0pt}{0pt}
 		$|V_{us}|$ & $0.224981$ 	\\
 		\rule[14pt]{0pt}{0pt}
 		$|V_{cb}|$ & $  0.0428095$ 	\\
 		\rule[14pt]{0pt}{0pt}
 		$|V_{ub}|$ & $0.00413679$ 	\\
 		\rule[14pt]{0pt}{0pt}
 		$\delta_{Q}$ & $61.3284^\circ$ 	\\
 		\rule[14pt]{0pt}{0pt}
 		$\chi^2$ & $10.6623$ 	\\
 		\hline
 	\end{tabular}
 	\caption{Benchmark point with common $\tau$ for quark sector.	 
 	}
 	\label{BP-q}
 \end{table}

 \begin{table}[t!]
 	\centering
 	\begin{tabular}{|c|c|} \hline 
 			\rule[14pt]{0pt}{0pt}
 		$\tau$&   $1.053 \, i$   \\ 
 		\rule[14pt]{0pt}{0pt}
 		$\epsilon_{M_1}$ &$ -1.34226 + 0.00413708 i$ \\
 		\rule[14pt]{0pt}{0pt}
 		$\epsilon_{M_2}$  &  $ 0.00945163 - 2.28048 i$ \\
 		\rule[14pt]{0pt}{0pt}
 		$\tilde y_{D_2}$ & $-0.0919822$  \\
 		\rule[14pt]{0pt}{0pt} 
 		$\tilde y_{D_3}$ &  $-0.324861$  \\
 		\rule[14pt]{0pt}{0pt}
		$\epsilon_{2}$ &  $-7.77596 - 0.0157862 i$  \\
 		\rule[14pt]{0pt}{0pt}
		$\epsilon_{3}$ &  $0.0415398 - 0.047376 i$  \\
 		\rule[14pt]{0pt}{0pt}
 		$\sin^2\theta_{12}$ & $0.320588$	\\
 		\rule[14pt]{0pt}{0pt}
 		$\sin^2\theta_{23}$ &  $ 0.457686$	\\
 		\rule[14pt]{0pt}{0pt}
 		$\sin^2\theta_{13}$ &  $0.0213432$	\\
 		\rule[14pt]{0pt}{0pt}
 		$\delta_{CP}^\ell$ &  $56.1248^\circ$ 	\\
 		\rule[14pt]{0pt}{0pt}
		$\alpha_{21},\ \alpha_{31}$ &  $241.826^\circ,\ 205.144^\circ $ 	\\
 		\rule[14pt]{0pt}{0pt}
 		$\sum m_i$ &  $57.4937$\,meV 	\\
 		\rule[14pt]{0pt}{0pt}
 		$\langle m_{ee} \rangle$ &  $2.53293$\,meV 	\\
 			\rule[14pt]{0pt}{0pt}
 		$\chi^2$ &  $5.62011$ 	\\
 		\hline
 	\end{tabular}
 	\caption{Benchmark point with common $\tau$ for lepton sector.	 }
 	\label{BP-l}
 \end{table}

\section{Dark matter and muon $g-2$}
{In this section we discuss the DM candidate in our model which has muon specific interaction due to the modular symmetry. 
Also muon $g-2$ is discussed considering the interaction.}

\subsection{Dark matter}
Here, we discuss DM candidate where we suppose it to be a CP-even real scalar DM $\chi\equiv \chi_{u_R}$
where $\chi_u\equiv (\chi_{u_R} + i \chi_{u_I})/\sqrt2$.
The relevant Lagrangian of DM including mass terms is given by 
\begin{align}
-{\cal L}_{DM} =  f_\mu \bar \mu E \chi_u -  \mu^2_\chi \chi_u^2 + y_E \bar E E \varphi_d 
+  |m_\chi|^2 |\chi_u|^2 +{\rm h.c.},
\end{align}
where $\mu^2_\chi$ arises from A-term $\varphi'_u \chi^2_u$ while $|m_\chi|^2$ comes from F-term of $\mu\chi_u\chi_d$ and soft mass B-term $|\chi_u|^2$.
Due to $\mu_\chi$, there is mass splitting between $\chi_{u_R}$ and $\chi_{u_I}$.
Thus, the mass of $E$ is given by VEV of $\varphi_d$; $m_E\equiv y_E v_{\varphi_d}/\sqrt2$ and the DM mass square is found as $|m_\chi|^2 - \mu^2_\chi$.
Our DM is real scalar and the dominant interaction is given by
\begin{align}
-{\cal L}_{DM}^{\rm int.} =  \frac{f_\mu}{\sqrt2} \bar \mu E \chi +{\rm h.c.}.
\end{align}
The dominant non-relativistic DM annihilation cross section is expanded in terms of relative velocity of DM $v_{\rm rel}$ and given by
\begin{align}
(\sigma v_{\rm rel})\approx \frac{|f_\mu|^4}{240\pi} \frac{m_\chi^6}{(m_\chi^2+ m_E^2)^4} v_{\rm rel}^4
\equiv d_{\rm eff} v_{\rm rel}^4 ,
\end{align}
where we suppose to be $m_\mu<<m_\chi\lesssim m_E$.
Then, the relic density is approximately written as~\cite{Bertone:2004pz}
\begin{align}
\Omega h^2\approx 
\frac{5.35\times 10^7 x_F^3{\rm GeV}^{-1}}{\sqrt{g_*(x_F)} M_{\rm PL} d_{\rm eff}}  , \label{eq:relic-rl}
\end{align}
where the present relic density is $0.1197$ at the best fit value~\cite{Planck:2013pxb}, $g_*(x_F\approx 25)\approx100$ counts the degrees of freedom for relativistic 
particles, and $M_{\rm PL}\approx 1.22\times 10^{19}$~GeV is the Planck mass.
From Eq.~(\ref{eq:relic-rl}), $f_\mu$ is written in terms of relic density as follows:
\begin{align}
|f_\mu|^4
& = 
\frac{5.35\times 10^7 x_F^3}{\sqrt{g_*(x_F)} M_{\rm PL}} \frac{240\pi(m_\chi^2+ m_E^2)^4}{m^6_\chi(\Omega h^2)\ {\rm GeV}} \nonumber \\
& \approx
(4.32\times10^{-5}) \times { \left( \frac{0.1197}{\Omega h^2} \right) } \times \frac{(m_\chi^2+ m_E^2)^4}{m^6_\chi\ {\rm GeV}^2}  
 ,
\label{eq:fmul}
\end{align}
where $|f_\mu|\lesssim \sqrt{4\pi}$ to satisfy perturbative limit.

\subsection{Muon anomalous magnetic dipole moment: $\Delta a_\mu$(muon $g-2$) \label{damu}}
\begin{figure}[tb]
\includegraphics[width=52.0mm]{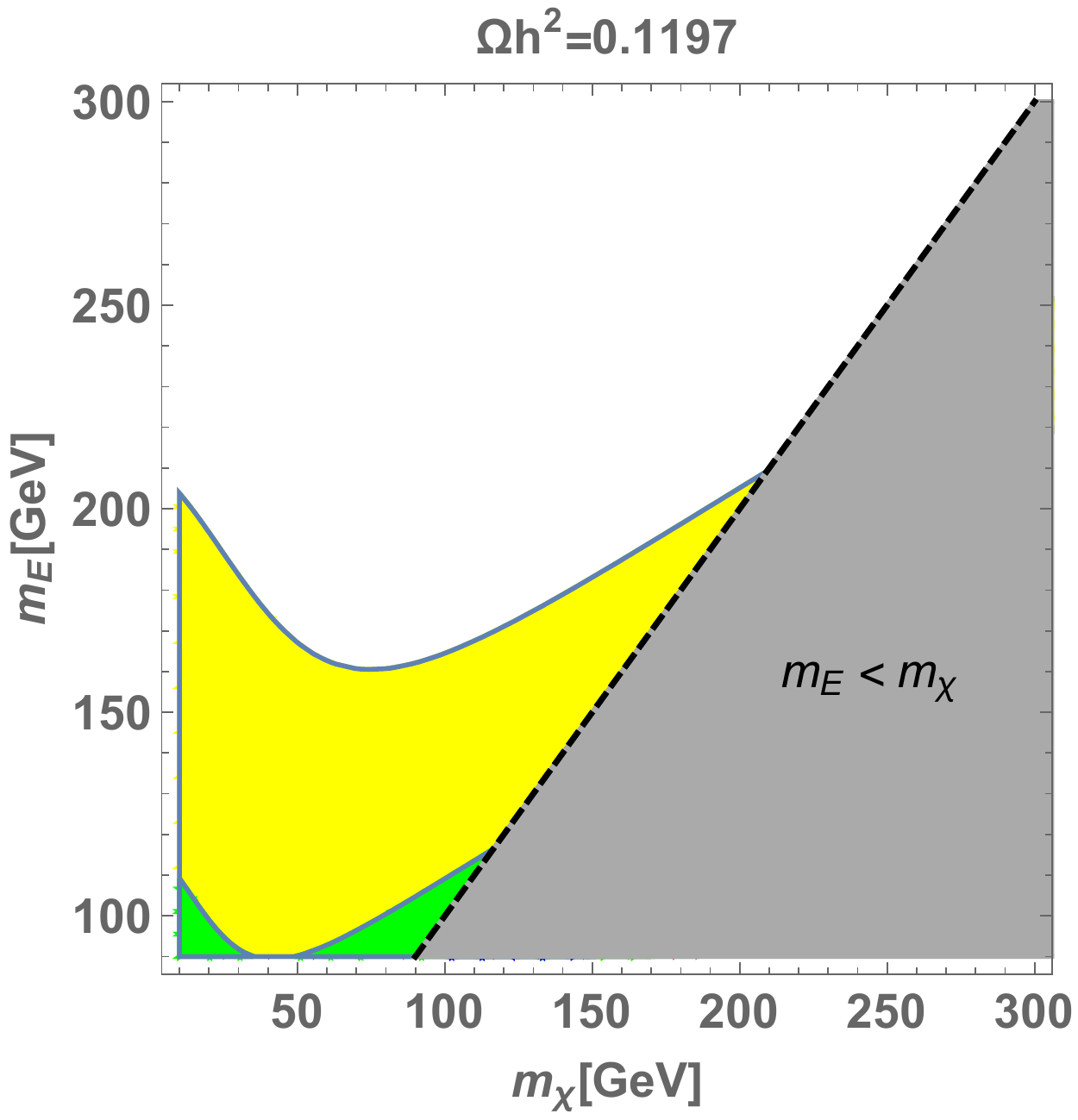}
\includegraphics[width=52.0mm]{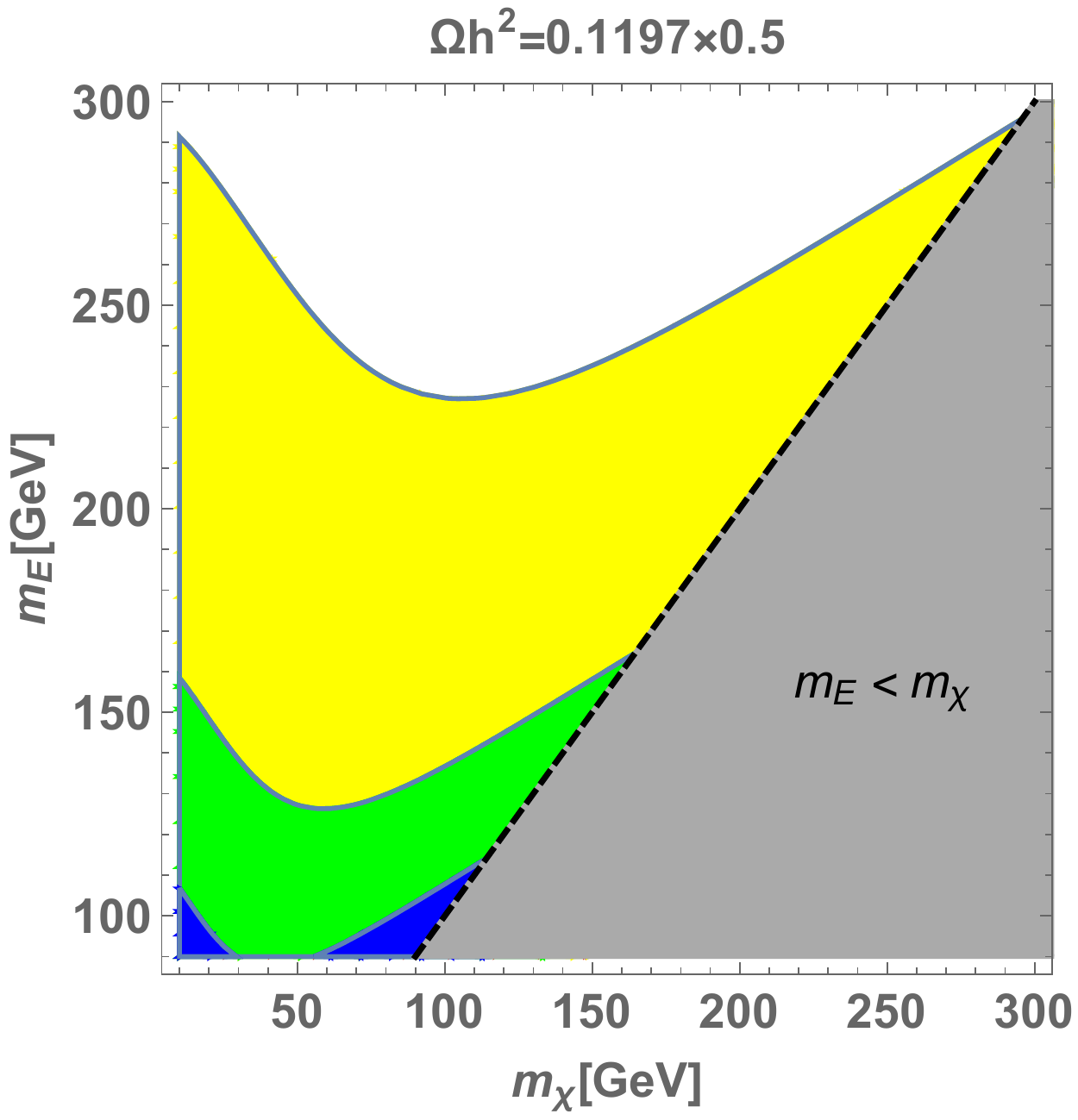}
\includegraphics[width=52.0mm]{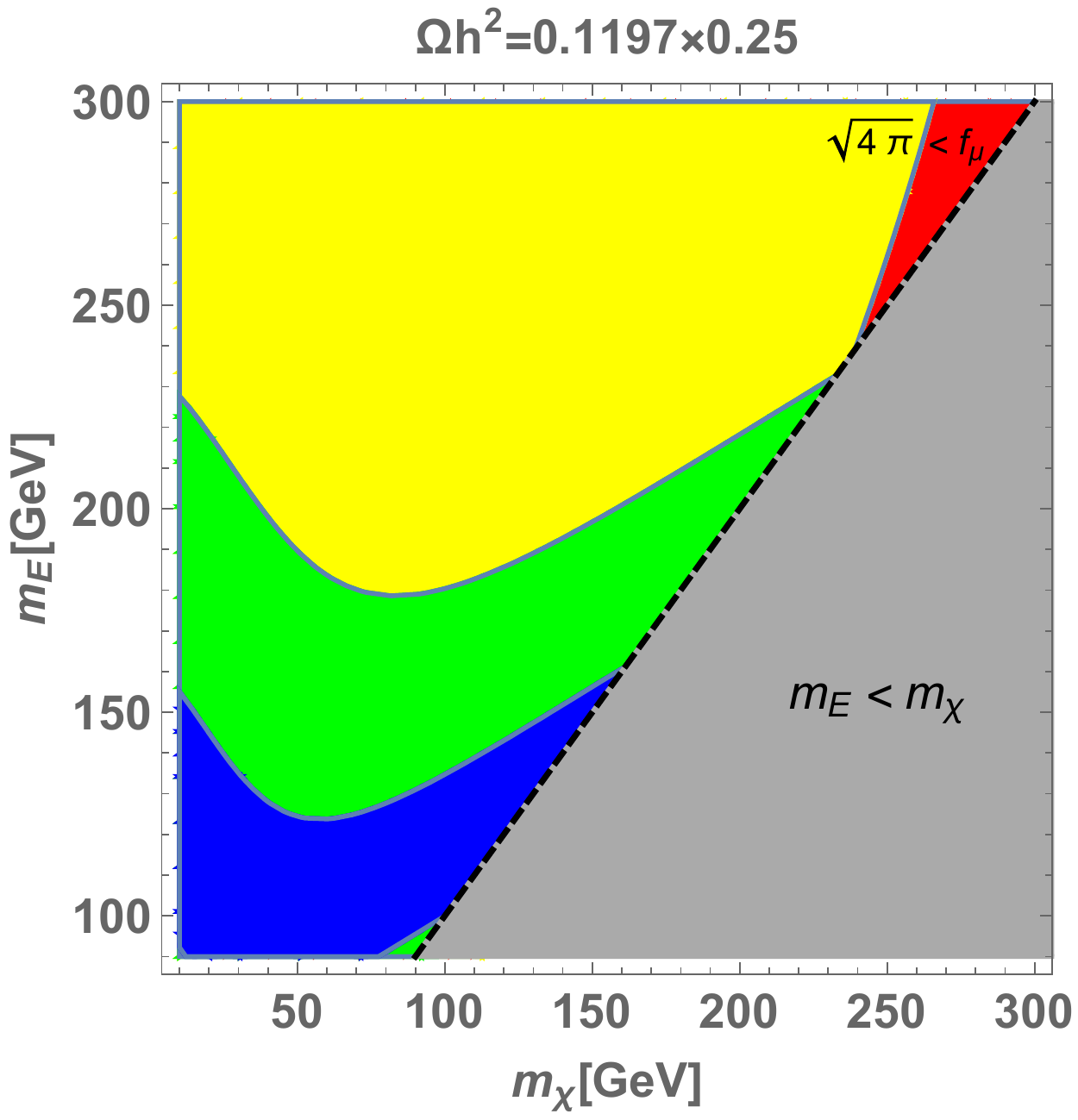}
 \caption{Allowed region between $m_\chi$ and $m_E$ in GeV unit with the relic density $\Omega h^2 = \{ 0.1197, 0.1197 \times 0.5, 0.1197 \times 0.25 \}$ for \{left, center, right\} plots, and satisfying muon $g-2$ where blue region is within $1\sigma$ interval of muon $g-2$,
 green $2\sigma$, and yellow $3\sigma$. Red space is excluded by the perturbative limit $f_\mu\lesssim\sqrt{4\pi}$.}
 \label{fig:dam-relic}
\end{figure}

The relevant Lagrangian for ${\Delta a_\mu}$ is the same as the one of DM and 
${\Delta a_\mu}$ is given by
\begin{align}
\Delta a_\mu 
=
\frac{|f_{\mu}|^2} {{16} \pi^2}
{\int_0^1 dx \frac{ m_\mu^2 x^2(1-x)}{x(x-1) m_\mu^2 + x m_E^2  + (1-x) m_\chi^2}} ~.
\label{eq:damu-res}
\end{align}
The current 4.2$\sigma$ deviation is given by~\cite{Hagiwara:2011af}~\footnote{{For a comprehensive review on new physics models for the muon $g-2$ anomaly as well as lepton flavor violation, see Ref.~\cite{Lindner:2016bgg}.}}
\begin{align}
\Delta a_\mu = (25.1\pm5.9)\times10^{-10} ~.
\label{eq:damu}
\end{align}
Inserting Eq.~(\ref{eq:fmul}) into Eq.~(\ref{eq:damu-res}), we rewrite the formula as follows:
\begin{align}
\Delta a_\mu 
=
\frac{(6.57\times10^{-3})}{16 \pi^2} \times {\left( \frac{0.1197}{\Omega h^2} \right)^{\frac12} } \times \frac{(m_\chi^2+ m_E^2)^2}{m^3_\chi\ {\rm GeV}}  
{\int_0^1 dx \frac{m_\mu^2 x^2(1-x)}{x(x-1) m_\mu^2 + x m_E^2  + (1-x) m_\chi^2}} ~.
\label{eq:damu-relic}
\end{align}
Here the Eq.~(\ref{eq:damu-relic}) is the formula taking the relic density into account.
In Fig.~\ref{fig:dam-relic}, we plot the allowed region between $m_\chi$ and $m_E$ in GeV unit applying Eq.~(\ref{eq:damu-relic}) 
{where we consider $\Omega h^2 = \{ 0.1197, 0.1197 \times 0.5, 0.1197 \times 0.25 \}$ for \{left, center, right\} plots}.
Here, the blue region is within $1\sigma$ interval of muon $g-2$,
 green one $2\sigma$, and yellow one $3\sigma$. Red space is excluded by the perturbative limit $f_\mu\lesssim\sqrt{4\pi}$ and the gray region is excluded since $E$ is stable there. 
{This figure suggests that the DM and $E$ mass is up to around 200 GeV within $3\sigma$ interval of muon $g-2$ when relic density is simultaneously satisfied. 
We also find that relic density will be smaller than the observed value when muon $g-2$ is within $1 \sigma$ interval. }

{
We also comment on phenomenology regarding hidden heavy fermions in the model.
The heavy charged lepton $E$ can be produced via electroweak interactions at the collider experiments such as the LHC since we require its mass is $\mathcal{O}(100)$ GeV scale to explain muon $g-2$ and relic density of the DM. 
The produced $E$ decays into muon and DM giving charged lepton plus missing transverse momentum signal which is similar to signal from slepton in a minimal supersymmetric model with neutralino being DM.
Comparing with constraints from slepton search in ref.~\cite{ATLAS:2018ojr}, a parameter region with mass difference 50 GeV $\lesssim m_E - m_\chi $ would be constrained and the region with small mass difference is favored.
Detailed analysis of collider constraint is beyond our scope of this work.
}

\section{Summary and Conclusions}
We have proposed a quark and lepton model explaining their masses, mixings, and CP violating phases, introducing a modular $A_4$ and hidden gauge $U(1)$ symmetries. Here, the hidden $U(1)$ accommodates heavier Majorana fermions that are required by chiral anomaly cancellations, and we have constructed a canonical seesaw scenario thanks to their neutral particles.
In this framework, we have searched for favorite parameter space to satisfy both the experimental values and would have obtained predictions, applying the $\chi$ square analysis.
Then, we have analyzed our real bosonic DM that annihilates into muon state only due to the modular $A_4$ flavor symmetry as well as muon $g-2$. Here, we do not need to consider any constraints of LFVs thanks to this flavor symmetry. Finally, we have shown the allowed space to satisfy the observed relic density of DM and muon $g-2$.


\section*{Acknowledgments}
The work of H.O. is supported by the Junior Research Group (JRG) Program at the Asia-Pacific Center for Theoretical
Physics (APCTP) through the Science and Technology Promotion Fund and Lottery Fund of the Korean Government and was supported by the Korean Local Governments-Gyeongsangbuk-do Province and Pohang City.
The work is also supported by the Fundamental Research Funds for the Central Universities (T.~N.).
H.O. is sincerely grateful for a lot of KIAS members.

\bibliography{MA4-QLDMmg2.bib}
\end{document}